\documentstyle[psfig,preprint,aps]{revtex}
\begin{document}
\tightenlines
\draft
\title{
{\begin{flushright}
{\footnotesize ZU-TH 25/01}\\
{\footnotesize YU-PP-I/E-KM-4-01}
\end{flushright}}
On the longitudinal contributions to
hadronic $\tau$ decay}
\author{Kim Maltman\thanks{e-mail: kmaltman@physics.adelaide.edu.au;
permanent address: Department of Mathematics and Statistics, 
York University, 4700 Keele St., Toronto, Ontario, CANADA M3J 1P3}}
\address{CSSM, University of Adelaide, Australia 5005 \\ and}
\address{Theory Group, TRIUMF, 4004 Wesbrook Mall, Vancouver, B.C.,
CANADA, V6T 2A3}
\author{Joachim Kambor\thanks{e-mail: kambor@physik.unizh.ch}}
\address{Institut f\"ur Theor. Physik, Univ. Z\"urich,
CH-8057 Z\"urich, Switzerland}
\maketitle
\begin{abstract}
A number of recent determinations of $m_s$ using hadronic $\tau$ decay
data involve inclusive analyses based on the so-called $(k,0)$
spectral weights.  We show here that the OPE
representations of the longitudinal contributions appearing
in these analyses, which are already known to be poorly converging,
have in addition an unphysical $k$ dependence which produces
a significant unphysical decrease in $m_s$ with increasing $k$.
We also show how, using additional sum rule constraints, 
the decay constants of the excited resonances in
the strange scalar and pseudoscalar channels may be
determined, allowing one to 
evaluate the longitudinal spectral contributions to the $(k,0)$
sum rules.  Taking into
account the very-accurately known $\pi$ and $K$ pole contributions,
we find that longitudinal contributions can be determined
with an accuracy at the few $\%$ level, and hence 
reliably subtracted, leaving an analysis for $m_s$ involving the sum of
longitudinal and transverse contributions, for which the OPE
representation is much better converged.
\end{abstract}
\pacs{12.15.Ff,11.55.Hx,12.38.Lg}

\section{Introduction}
The determination of the strange quark mass, $m_s$, 
has been the focus of much recent activity in both the sum
rule~\cite{srss,jm,sr38,srsps,ck93,kmtpr,pp98,cdh,ckp98,pp99,kkp00,km00,D00,C00}
and lattice~\cite{lat1,lat2,lat3,latticerecent} communities.
(A summary of the current status of both the sum rule and
lattice determinations
is given in Ref.~\cite{gm01rev}.)

Among the sum rule approaches, those based on flavor breaking in
hadronic $\tau$ decay~\cite{ck93,kmtpr,pp98,cdh,ckp98,pp99,kkp00,km00,D00,C00}
appear the most reliable at present.  There are two reasons for this
statement.  First, on the experimental side,
the spectral data required is known over the full kinematic range
entering the relevant sum rules~\cite{ALEPHud,ALEPHstrange} and, second,
on the theoretical (OPE) side, flavor-independent
instanton and renormalon effects, which create 
potential uncertainties in
analyses of the strange scalar and pseudoscalar channels~\cite{srss,jm,srsps}, 
cancel in forming the flavor-breaking $\tau$ decay 
difference~\cite{pp98}.  The $\tau$ decay sum rules 
are, however, not without complications.  The primary complication
has to do with the behavior of the OPE representation of the
contributions to the inclusive decay rate 
of hadronic states with total spin $J=0$.
In order to elaborate on this point,
and to fix terminology and notation, 
we briefly review the relation of $m_s$ to hadronic $\tau$
decay.

The ratio of the inclusive hadronic $\tau$ decay width
through the flavor $ij=ud$ or $us$ vector (V) or axial vector (A) 
weak hadronic current
to the corresponding electron
decay width,
\begin{equation}
R_{ij;V,A} \equiv \frac{ \Gamma [\tau^- \rightarrow \nu_\tau
\, {\rm hadrons_{ij;V,A}}\, (\gamma)]}{ \Gamma [\tau^- \rightarrow
\nu_\tau e^- {\bar \nu}_e (\gamma)] } ,
\label{one}\end{equation}
where $(\gamma )$ indicates additional photons or lepton pairs,
can be expressed
in terms of weighted integrals over the spin $J=0$
(longitudinal) and $J=1$ (transverse) components
of the corresponding V or A spectral functions~\cite{tau1}, where
the spectral functions are defined, as usual, by
\begin{equation}
\rho^{(J)}_{ij;V,A}(q^2)\equiv {\frac{1}{\pi}}\, {\rm Im}\, 
\Pi^{(J)}_{ij;V,A}(q^2)\ .
\label{pispin}\end{equation}
In Eq.~(\ref{pispin}), $\Pi^{(J)}_{ij;V,A}$ are the spin 
$J$ scalar components of the
usual flavor $ij$ V, A current-current correlators, 
\begin{eqnarray}
i \int d^4x \, e^{iq\cdot x} &&
\langle 0\vert T\Bigl(
J^{\mu}_{ij;V,A}(x) J^{\nu}_{ij;V,A}(0)^\dagger 
\Bigr)\vert 0\rangle \, \nonumber \\
&&\qquad\qquad\equiv \left( -g^{\mu\nu} q^2 + q^{\mu} q^{\nu}\right) \, 
\Pi_{ij;V,A}^{(1)}(q^2)
 + q^{\mu} q^{\nu} \, \Pi_{ij;V,A}^{(0)}(q^2)\, .
\end{eqnarray}
Working with the combinations
$\Pi^{(0+1)}(q^2)\equiv \Pi^{(0)}(q^2)+\Pi^{(1)}(q^2)$ and
$q^2\, \Pi^{(0)}(q^2)$ which have no kinematic singularities,
$R_{ij;V,A}$ can then be written~\cite{tau2,bnp}
\begin{eqnarray}
R_{ij;V,A} &=&  
12 \pi^2 S_{EW}\, \vert V_{ij}\vert^2 
\int^{m_\tau^2}_0 {\frac{ds} {m_\tau^2 }} \,
\left( 1-{\frac{s}{m_\tau^2}}\right)^2 
\left[ \left( 1 + 2 {\frac{s}{m_\tau^2}}\right) 
\, \rho^{(0+1)}_{ij;V,A}(s) 
-{\frac{2s}{m_\tau^2}} \, \rho^{(0)}_{ij;V,A}(s) \right] 
\nonumber \\
&=& 6 \pi S_{EW}\,  \vert V_{ij}\vert^2 
i\, \oint_{|s|=m_\tau^2} {\frac{ds}{ m_\tau^2}}\,
\left( 1- {\frac{s}{ m_\tau^2}}\right)^2 \left[ \left( 
1 + 2 {\frac{s}{m_\tau^2}}\right) \Pi^{(0+1)}_{ij;V,A}(s)
- 2 {\frac{s}{ m_\tau^2}}\, \Pi^{(0)}_{ij;V,A}(s) \right] \, ,
\label{taufesr}
\end{eqnarray}
where $s=q^2=-Q^2$, $S_{EW}=1.0194$ represents the leading electroweak 
corrections\cite{ms88}, $V_{ij}$ are the usual CKM matrix elements,
and the second line follows from the first as a consequence of Cauchy's
theorem.  The second line of Eqs.~(\ref{taufesr})
allows $R_{ij;V,A}$ to be evaluated using techniques based on the OPE
and perturbative QCD~\cite{tau1,tau2,bnp,ledp92,pichrev}.
The weights, $w_T(y)\equiv \left(1-y\right)^2\left( 1+2y\right)$
and $w_L(y)=-2y\left(1-y\right)^2$ (with $y\equiv s/m_\tau^2$)
multiplying $\Pi^{(0+1)}$ and
$\Pi^{(0)}$, respectively, have (double) zeros 
at $s=m_\tau^2$, reflecting the
fact that $s=m_\tau^2$ lies at the edge of hadronic phase
space.  The resulting suppression of contributions from the
region of the circular contour near the timelike real
axis (the region of potential breakdown of the OPE)
is responsible for the high quality of the OPE representation
of the inclusive hadronic rates (see Ref.~\cite{pichrev} for a
recent review).

Defining $R_{ij}\equiv R_{ij;V}+R_{ij;A}$, it is then evident that
\begin{equation}
\Delta R_\tau \equiv {\frac{R_{ud}}{\vert V_{ud}\vert^2}}
- {\frac{R_{us}}{\vert V_{us}\vert^2}} 
\end{equation}
vanishes in the $SU(3)_F$ limit.  Defining
$\Delta\Pi^{(J)}\equiv \Pi^{(J)}_{ud}-\Pi^{(J)}_{us}$ and
$\Delta\rho^{(J)}\equiv\rho^{(J)}_{ud}-\rho^{(J)}_{us}$,
the mass-independent $D=0$ contributions, therefore, cancel
by construction on the OPE side of the sum rule for $\Delta R_\tau$
analogous to Eq.~(\ref{taufesr}),
\begin{eqnarray}
&&12\pi^2\, S_{EW}\, \int^1_0 {dy} \,
\left( w_T(y) 
\Delta\rho^{(0+1)}(s)\, + \, w_L(y)\Delta\rho^{(0)}(s)
\right) \nonumber \\
&&\qquad\qquad\qquad\qquad =
6\pi {\rm i}\, S_{EW}\, \oint_{\vert y\vert =1}\, dy\,
\left( w_T(y)\Delta\Pi^{(0+1)}(s)+w_L(y)
\Delta\Pi^{(0)}(s)\right)\ .
\label{fbfesr}
\end{eqnarray}
Neglecting $m^2_{u,d}$ and $\alpha_s m_{u,d}m_s$ 
relative to $m_s^2$, the leading $D=2$ term 
is proportional to $m_s^2$\cite{ck93,bnp}.
The integrand on the LHS of Eq.~(\ref{fbfesr})
is known, as a function of $s$, from the work of the 
ALEPH collaboration~\cite{ALEPHud,ALEPHstrange}.
On the OPE side, the $D=2$ contribution is known
in terms of $\alpha_s$, once $m_s$ is fixed; the
$D=4$ contribution is known in terms of $\langle m_s \bar{s}s\rangle$;
and the $D=6$ contribution is small as a result of the
cancellation between the dominant $D=6$ four-quark condensate
terms which occurs in the V+A sum~\cite{ck93,bnp}.  Alternate sum rules which
also allow the spectral side to be evaluated using 
the measured $s$-dependent $ud$ and
$us$ number distributions, without necessitating
a $J=0/J=1$ separation of the experimental data, can
be constructed by multiplying the integrands 
appearing on both sides of Eq.~(\ref{fbfesr}) by a
common analytic factor.  For the case that this factor is $(1-y)^ky^n$, 
the resulting sum rule is said to involve the $(k,n)$
spectral weight, and the resulting analogue of
$\Delta R_\tau$ is denoted $\Delta R^{(k,n)}_\tau$.  
The $(k,0)$ spectral weight
sum rules (with $k=0,1,2$) form the basis of
a number of recent inclusive treatments of the $m_s$ extraction 
problem~\cite{pp98,cdh,ckp98,pp99,kkp00,D00,C00}.
We will denote the weights accompanying $\Pi^{(0+1)}$
and $\Pi^{(0)}$ in the $(k,0)$ sum rules by
$w_T^{(k)}(y)=\left( 1-y\right)^{2+k}
\left( 1+2y\right)$ and $w_L^{(k)}(y)=-2y\left( 1-y\right)^{2+k}$,
respectively.

Eq.~(\ref{fbfesr}) (corresponding to the $(0,0)$ spectral
weight), and/or its $(k,0)$ generalizations, would 
allow a straightforward determination of
$m_s$, provided the OPE representations
of both the $J=0+1$ and $J=0$ contributions above were
well converged at scale $m_\tau^2$.  
Unfortunately, it turns out that this is not the case.
The problem lies with the OPE contribution
involving the product $w^{(k)}_L(y)\Pi^{(0)}(s)$.  
We refer to these contributions as ``longitudinal'' in what follows.

The source of the problem is that the perturbative series for the 
integrated $D=2$ longitudinal contribution
(which is known to four loops, i.e.,
$O(\alpha_s^3)$~\cite{chetyrkin97,cps})
is not convergent at the scale
$m_\tau^2$~\cite{kmtpr,pp98,kkp001}.
This is true whether one considers the
``fixed order'' (FOPT) expansion
(expansion in powers of $\alpha_s(\mu^2)$ at a fixed scale, $\mu$,
e.g., $\mu =m_\tau$), or the ``contour-improved''
expansion (CIPT)~\cite{ledp92,pivcipt} (in which the large logarithms
are summed up by the scale choice $\mu^2=Q^2$ point-by-point
over the contour).  Taking the unmodified version of
Eq.~(\ref{fbfesr}), corresponding to $k=0$, to be specific, one finds that,
using the central value of the ALEPH determination,
$\alpha_s(m_\tau^2)=.334$, the FOPT $D=2$ series behaves as
\begin{equation}
\sim 1+0.99+1.24+1.59+\cdots \ ,
\end{equation}
while the corresponding CIPT series behaves as~\cite{ALEPHstrange}
\begin{equation}
\sim 1+0.78+0.78+0.90+\cdots
\end{equation}
(where in both cases we have normalized successive terms in
the integrated series to the leading, $O(\alpha_s^0)$, term).
The integrated longitudinal $D=2$ series, truncated at
$O(\alpha_s^3)$, also exhibits a very strong residual scale
dependence~\cite{pp98}.  Because of the non-convergence and
strong residual scale dependence, inclusive 
sum rules of the type described above 
contain significant uncertainties associated with the presence
of the longitudinal contributions.  Recent inclusive
analyses~\cite{pp98,ckp98,pp99,kkp00,D00,C00,ALEPHstrange} proceed by
taking the sum of $D=2$ $J=0+1$ and $J=0$ contributions 
to $O(\alpha_s^3)$.  The size of the last known
($O(\alpha_s^3)$) term is taken as an estimate of 
the $D=2$ truncation error.
In Refs.~\cite{pp98,pp99,D00,C00},
an additional error has been included to account for the
residual scale dependence of the truncated sum.  The scale-dependence
error is estimated by varying the scale choice in the CIPT evaluation
of the $D=2$ sum according to $\mu^2 =\xi^2Q^2$, with $.75<\xi <2$,
and symmetrizing the resulting variation about the central
value $\xi =1$.  The resulting estimated $D=2$ error is much
larger than that for the remaining longitudinal terms,
and hence dominates the error on the total longitudinal 
contribution.  Taking the results of Ref.~\cite{pp99},
which gives the detailed breakdown into individual longitudinal 
contributions, to be specific, and combining the quoted errors
in quadrature, one finds errors of $32\%$, $34\%$
and $37\%$ for the $(0,0)$, $(1,0)$ and $(2,0)$ total longitudinal
contributions, respectively.  These errors represent
the dominant component of the total theoretical error in
the inclusive analyses.  

While the errors on the longitudinal contributions
discussed above might appear safely conservative, we will argue in this
paper that they are, in fact, almost certainly too small.  We
will, in addition, demonstrate that the central values 
of the OPE contribution to $\Delta R_\tau^{(k,0)}$ contain
an unphysical dependence on $k$ which produces a corresponding
unphysical lowering of the extracted value of $m_s$ with 
increasing $k$.  Finally, we will demonstrate that it is possible
to significantly reduce the uncertainties on the total longitudinal
contributions, allowing one to subtract the longitudinal
contributions from the experimental number distribution and
work instead with sum rules for the much better behaved
$0+1$ correlator difference.  

The rest of the paper is organized
as follows.  In Section II we demonstrate the existence
of the problem and investigate its magnitude.
In Section III we discuss how to improve the estimate
for the longitudinal contributions using sum rules for the
strange scalar and pseudoscalar channels and present our numerical results.
Finally, in Section IV, we summarize our results and comment
on their implications for future analyses of $m_s$ using
$\tau$ decay data.

\section{Physical Constraints on the Longitudinal Contributions
to the Spectral Weight Analyses}

If a $J=0/J=1$ spin decomposition existed for the current experimental
data, one could simply subtract the
longitudinal contribution from each bin of the experimental 
$ud$ and $us$ distributions,
determine $\rho^{(0+1)}_{ud,us;V+A}(s)$,
and use this information to analyze the 
$0+1$, rather than inclusive, sum rules.  Unfortunately, such a 
decomposition does not yet exist over the whole of the kinematically
allowed range.  Certain general features of the longitudinal contributions to
the V and A correlators, however, allow us to, nonetheless, 
obtain useful constraints.

In the chiral limit, the longitudinal spectral
functions vanish except for the (massless) $\pi$ and $K$ pole 
terms, which contribute to
$\rho^{(0)}_{ud;A}$ and $\rho^{(0)}_{us;A}$, respectively.
Away from the chiral limit, 
$\rho^{(0)}_{ij;V}$ and
$\rho^{(0)}_{ij;A}$ receive additional contributions 
proportional, respectively, to $(m_i- m_j)^2$ and $(m_i+m_j)^2$.
For $ij=ud$, these additional contributions are
numerically tiny and can be neglected.  For $ij=us$,
the thresholds for the non-pole contributions to $\rho^{(0)}_{us;V}$ and
$\rho^{(0)}_{us;A}$ are $(m_K+m_\pi )^2\equiv s_{th}^{SS}$ and
$(m_K+2m_\pi )^2\equiv s_{th}^{SPS}$, respectively.  The sum of
the longitudinal $K$ and $\pi$ pole contributions
for the $(k,0)$ spectral weight is
\begin{equation}
\left[ \Delta R^{(k,0)}_\tau\right]^{K+\pi}_L = 48\pi^2\, S_{EW}\left[
\left( {\frac{f_K^2}{m_\tau^2}}\right) \left( {\frac{m_K^2}{m_\tau^2}}
\right) \left( 1-{\frac{m_K^2}{m_\tau^2}}\right)^{2+k}
- \left( {\frac{f_\pi^2}{m_\tau^2}}\right) \left( {\frac{m_\pi^2}{m_\tau^2}}
\right) \left( 1-{\frac{m_\pi^2}{m_\tau^2}}\right)^{2+k}\right]\ ,
\label{longpole}
\end{equation}
with $f_\pi =92.4\ {\rm MeV}$ and $f_K =113.0\ {\rm MeV}$~\cite{pdg2000}.
The $\pi$ pole contribution is nearly constant with increasing
$k$, the $K$ pole contribution slowly decreasing with $k$.
The remaining longitudinal contribution, which we will refer
to as the ``resonance contribution'', is then given by
\begin{equation}
\left[ \Delta R^{(k,0)}_\tau\right]^{resonance}_L 
= 12\pi^2\, S_{EW}\, \int_{y_{th}}^1\, dy\ 2y\left( 1-y\right)^{2+k}
\left[ \rho^{(0)}_{us;V}+\rho^{(0)}_{us;A}\right]\ ,
\label{longres}
\end{equation}
where $y_{th}=s_{th}^{SS}/m_\tau^2$.
The V part of this contribution should be dominated by
the $K_0^*(1430)$ resonance, since only the tail of the next strange scalar
resonance, the $K_0^*(1950)$, lies within the kinematically allowed range,
$s< m_\tau^2$. Similarly, the A contribution in Eq.~(\ref{longres})
should be dominated by the $K(1460)$.
The $K$ and $\pi$ pole longitudinal contributions are, of course,
very accurately known, so it is the absence of an experimental
determination of the $K_0^*(1430)$ and $K(1460)$ decay constants
which prevents us from performing a reliable longitudinal subtraction.

Note that, because $\rho^{(0)}_{us;V}$ and $\rho^{(0)}_{us;A}$ are
positive, and the analogous $ud$ resonance contributions 
are negligible, the longitudinal resonance contribution 
of Eq.~(\ref{longres}) is necessarily positive and a decreasing
function of $k$.  In fact, from Eq.~(\ref{longres}) it follows
that the longitudinal $(k,0)$
resonance contributions must satisfy the inequality
\begin{equation}
\left[ \Delta R^{(k+1,0)}_\tau\right]^{resonance}_L \leq
\left( 1-{\frac{s_{th}^{SS}}{m_\tau^2}}\right)
\left[ \Delta R^{(k,0)}_\tau\right]^{resonance}_L
=0.873\left[ \Delta R^{(k,0)}_\tau\right]^{resonance}_L\ ,
\label{minconstraint}
\end{equation}
where the equality would obtain only if the entire non-pole $us$
spectral strength lay at threshold in the scalar channel.
That both the pole and resonance contributions
are decreasing functions of $k$, of course, also means
that the {\it total} longitudinal contribution
to $\Delta R^{(k,0)}_\tau$ must be a decreasing function of $k$.

The fact that one expects only very small contributions from the
tails of the $K_0^*(1950)$ and $K(1830)$ resonances, and that
both the masses and widths of the $K_0^*(1430)$ and $K(1460)$
are similar allows us to sharpen considerably
the constraint represented by Eq.~(\ref{minconstraint}).  
In the narrow width approximation (NWA),
dominance by resonance contributions at $M\sim 1.4\ {\rm GeV}$
would mean that   
\begin{equation}
\left[ \Delta R^{(k+1,0)}_\tau\right]^{resonance}_L \simeq
\left( 1-{\frac{M^2}{m_\tau^2}}\right)
\left[ \Delta R^{(k,0)}_\tau\right]^{resonance}_L
=0.38\, \left[ \Delta R^{(k,0)}_\tau\right]^{resonance}_L\ .
\label{nwaconstraint}
\end{equation}
A more refined version of this estimate is obtained by
considering $K_0^*(1430)$ and $K(1460)$ Breit-Wigner forms
with PDG2000 values for the masses and widths, and integrating
directly over the resonance profiles.  The results of this
exercise are that the individual resonance contributions to 
$\left[ \Delta R^{(k,0)}_\tau\right]_L$ 
for $k=0,1,2$ are in the ratios $1:0.46:0.24$ for the
$K_0^*(1430)$ and $1:0.43:0.20$ for the $K(1460)$.
We would thus expect the total resonance contributions
to $\left[ \Delta R^{(k,0)}_\tau\right]_L$ to satisfy
\begin{eqnarray}
\left[ \Delta R^{(1,0)}_\tau\right]^{resonance}_L &\simeq&
0.44\, \left[ \Delta R^{(0,0)}_\tau\right]^{resonance}_L
\nonumber \\
\left[ \Delta R^{(2,0)}_\tau\right]^{resonance}_L &\simeq&
0.22\, \left[ \Delta R^{(0,0)}_\tau\right]^{resonance}_L\, .
\label{physicalconstraint}
\end{eqnarray}

Let us now consider the OPE representation of
$\left[ \Delta R^{(k,0)}_\tau\right]_L$.
In what follows, we will, for convenience,
quote the OPE results obtained in Ref.~\cite{pp99}, 
since that reference provides a complete breakdown of the individual
$0+1$ and longitudinal contributions (in addition to a 
detailed discussion of the evaluation of the various contributions, 
to which the interested reader is referred for details).
The $D=2$ longitudinal contributions are of the form~\cite{pp99}
\begin{equation}
\left[ \Delta R^{(k,0)}_\tau\right]^{D=2}_L =
6\, S_{EW}\left( 1-\epsilon_d^2\right)\left({\frac{m_s(m_\tau^2)}
{m_\tau^2}}\right)^2\, \Delta^L_{(k,0)}\ ,
\label{longd2}
\end{equation}
where $\epsilon_d=m_d/m_s=.053$, and $\Delta^L_{(k,0)}$, which
results from the CIPT integration, depends on $\alpha_s(m_\tau^2)$.
The results of Ref.~\cite{pp99} are
\begin{eqnarray}
\Delta^L_{(0,0)} &=& 5.1\pm 2.1\pm 0.5\nonumber \\
\Delta^L_{(1,0)} &=& 5.3\pm 2.5\pm 0.7\nonumber \\
\Delta^L_{(2,0)} &=& 5.8\pm 3.2\pm 0.8\ ,
\label{pp99d2values}
\end{eqnarray}
where, in each case, the first error represents the combination
of the scale-dependence and truncation errors, discussed above,
and the second represents the effect 
of the experimental uncertainty in
$\alpha_s(m_\tau^2)$.
The $D=4$ longitudinal contributions are, numerically, 
completely dominated
by the term proportional to $\langle m_s \bar{s}s\rangle$.  Since
this term arises from the Ward identity, 
\begin{equation}
q^4\, \Pi^{(0)}_{ij;V,A}(q^2)=(m_i\pm m_j)^2\, \Pi_{ij;S,P}(q^2)
+(m_i\pm m_j)\left(\langle \bar{q}_iq_i\rangle
\pm \langle \bar{q}_jq_j\rangle\right)\ ,
\label{wardID}
\end{equation}
where $\Pi_{ij:S,P}$ are the correlators of the flavor
$ij$ scalar and pseudoscalar densities, and 
the plus (minus) signs on the RHS correspond to the A (V) case,
the Wilson coefficient of the $\langle m_s \bar{s}s\rangle$ term
receives no radiative corrections.  The $D=4$ 
contribution can therefore be evaluated rather accurately,
using (i) the quark mass ratios obtained by
Leutwyler~\cite{leutwylermq}, (ii) the GMOR relation,
$\langle (m_u+m_d)\bar{u}u\rangle = -f_\pi^2 m_\pi^2$,
{\begin{footnote}{Deviations from the GMOR relation have recently been shown 
to be at most $6\%$ \cite{CGL01}. The resulting error on the 
$m_s$ analysis is completely negligible.}\end{footnote}}
and (iii) $\langle \bar{s}s\rangle /\langle \bar{u}u\rangle 
= 0.8\pm 0.2$~\cite{pp99}.
The value turns out to be the same for the $(0,0)$, $(1,0)$, and
$(2,0)$ spectral weights~\cite{pp99}:
\begin{equation}
\left[ \Delta R^{(k,0)}_\tau\right]^{D=4}_L = 0.0726\pm 0.0194\ .
\label{longd4}
\end{equation}
The error in Eq.~(\ref{longd4}) is considerably
smaller than that on the $D=2$ contributions.  The leading
four-quark $D=6$ contributions are absent from 
$\Pi^{(0)}_{ij;V,A}$~\cite{bnp},
justifying the neglect of $D=6$ contributions, while
contributions of $D=8$ and higher are assumed to be negligible.
The OPE representation of $\left[ \Delta R^{(k,0)}_\tau\right]_L$
employed in recent inclusive analyses
thus consists of the sum of $D=2$ and
$D=4$ terms, with the error dominated by that on the
truncated $D=2$ series.

It is now straightforward to see that the OPE 
representation just described for $\left[ \Delta R^{(k,0)}_\tau\right]_L$
rather badly violates the constraints obtained above.  
We first observe that, from Eqs.~(\ref{longd2}) and 
(\ref{pp99d2values}),
$\left[ \Delta R^{(k,0)}_\tau\right]^{D=2}_L$ is a slowly
increasing function of $k$.
Since $\left[ \Delta R^{(k,0)}_\tau\right]^{D=4}_L$ is constant
with $k$, this means that
the OPE representation,
$\left[ \Delta R^{(k,0)}_\tau\right]^{OPE}_L$ is also slowly increasing, rather
than decreasing, with $k$.{\begin{footnote}{The small 
$O(m_s^4)$ contributions
to $\left[ \Delta R^{(k,0)}_\tau\right]^{D=4}_L$, which have
been neglected above, are actually also increasing with $k$,
so the full longitudinal $D=4$ contribution is actually also (very
slowly) increasing with $k$.}\end{footnote}}
Since, for values of 
$m_s(m_\tau^2)$ typical of those obtained in recent sum rule
analyses, $m_s(m_\tau^2)\sim 120\ {\rm MeV}$, the $D=2$
contributions are more than a factor of $2$ larger than the
corresponding $D=4$ contribution, this problem 
is likely to have important numerical consequences.
The increase with $k$ of the central values shown in
Eq.~(\ref{pp99d2values}), in fact, 
means that the extracted values of $m_s$ must {\it necessarily}
display an unphysical {\it decrease} with $k$.
{\begin{footnote}{It is worth noting that Ref.~\cite{C00}
employs a $k$-dependent truncation scheme, in
contrast to the other inclusive analyses, which truncate
at the same order for all $k$.
The problem of the unphysical $k$-dependence of the
central values outlined above, however,
remains present even for this altered
scheme.}\end{footnote}}
This unphysical decrease would not, of course, be
a practical (as opposed to conceptual) difficulty
for the inclusive analysis if $\left[ \Delta R^{(k,0)}_\tau\right]^{OPE}_L$ 
represented only a small fraction of $\Delta R_\tau$.
Unfortunately, this is not the case.  To illustrate
this point we show, in Table~\ref{table1}, for each of the three
$(k,0)$ spectral weights, the numerical values
of the longitudinal $D=2$ contribution and total longitudinal
contribution obtained in Ref.~\cite{pp99}, together with the 
1999 ALEPH experimental values of $\Delta R^{(k,0)}_\tau$~\cite{ALEPHstrange}.
The reader should be aware that significantly
different central values of $m_s(m_\tau^2)$ were obtained
for the three different spectral weights; the tabulated
$D=2$ contributions correspond to these central values
(also listed in the table).  The quoted $D=2$ errors were obtained
by combining the two theoretical errors in 
Eq.~(\ref{pp99d2values}) in quadrature.  We see from the table
that the longitudinal contribution, in each case, represents
more than half of the experimental value.

Let us attempt to quantify how large the
errors associated with the unphysical $k$ dependence of
$\left[ \Delta R^{(k,0)}_\tau\right]^{OPE}_L$
might be.  Note that, given the accurately known
values of the longitudinal $\pi$ and $K$ pole contributions, the
OPE representation of $\left[ \Delta R^{(k,0)}_\tau\right]_L$ should be
thought of as providing an estimate for the sum of
the unknown longitudinal resonance contributions.  
In Table~\ref{table2} we display the values of
$\left[ \Delta R^{(k,0)}_\tau\right]^{resonance}_L$
implicit in the OPE results of Ref.~\cite{pp99}, together
with the values of the corresponding pole contributions,
$\left[ \Delta R^{(k,0)}_\tau\right]^{K+\pi}_L$.
Two versions of the resonance contributions are given.
In the first (labelled ``IND'' in the table), 
the values of $m_s(m_\tau^2)$
used to evaluate the three different
$\left[ \Delta R^{(k,0)}_\tau\right]^{OPE}_L$ are different,
corresponding to the central values obtained in the relevant 
independent $(k,0)$ analysis of Ref.~\cite{pp99}.
In the second (labelled ``COMB'' in the table),
the common value, $m_s(m_\tau^2)=119\ {\rm MeV}$, obtained
in the combined fit to all three $(k,0)$ sum rules~\cite{pp99}, is used
to compute all three of the $\left[ \Delta R^{(k,0)}_\tau\right]^{OPE}_L$.
The unphysical increase of $\left[ \Delta R^{(k,0)}_\tau\right]^{OPE}_L$
with $k$, combined
with the decrease of the sum of the $\pi$ and $K$ pole contributions
with $k$, means that the nominal resonance contribution implicit
in $\left[ \Delta R^{(k,0)}_\tau\right]^{OPE}_L$
must be increasing with $k$, for fixed
$m_s(m_\tau^2)$.  This is evident in the results of the 
combined analysis, but
obscured by the decrease of $m_s(m_\tau^2)$ with $k$ for the
independent analysis.  It is evident that both sets of
results are far from satisfying the constraints given
in Eqs.~(\ref{physicalconstraint}).  

We are now in a position to illustrate the potential significance
of the unphysical $k$ dependence of 
$\left[ \Delta R^{(k,0)}_\tau\right]^{OPE}_L$
on the extracted values of $m_s$.  Let us imagine that
the central OPE value provides a good
approximation for one of the three 
$\left[ \Delta R^{(k,0)}_\tau\right]^{resonance}_L$, and use
Eq.~(\ref{physicalconstraint}) to estimate the resonance
contributions to the other two $(k,0)$ sum rules.  We 
find that, if we attempt to make this assumption for 
either $k=1$ or $k=2$ the result is a
$k=0$ longitudinal contribution which exceeds the full experimental
value, violating the positivity of the $0+1$ OPE
representation.  If we instead make the assumption for $k=0$,
the resulting change in the $(1,0)$ and $(2,0)$
longitudinal contributions produces a shift in
the extracted central $m_s(m_\tau^2)$ values from
$121\rightarrow 142\ {\rm MeV}$ and $106\rightarrow 133\ {\rm MeV}$,
respectively.  We stress that this exercise is for illustrative purposes
only; although the consistency of the three 
analyses is significantly improved if one assumes that,
for some reason, the $k=0$ representation is good,
the resulting
``extraction'' of $m_s$ is meaningless since
the assumption simultaneously forces the $k=1,2$ representations
to be bad, leading one to the conclusion that it is
unreasonable to have assumed that the $k=0$ representation
was good in the first place.

At present there is little experimental information available
on the size of the longitudinal resonance contributions.  
The PDG2000 compilation
provides no information on $\tau\rightarrow K(1460)\nu_\tau$,
and quotes only an upper
bound, $B< .0005$ on the $\tau\rightarrow K_0^*(1430)\nu_\tau$ 
branching fraction.  The latter bound corresponds to an upper bound 
$\left[ \Delta R^{(0,0)}_\tau\right]^{K_0^*(1430)}_L< 0.052$.
The central longitudinal $(0,0)$ OPE determination, 
if reliable, would then
require a corresponding $K(1460)$ contribution {\it greater}
than $\sim 0.10$.  Taking a Breit-Wigner $K(1460)$ form 
with PDG2000 values for the mass and width, this corresponds to
$f_{K(1460)}>100\ {\rm MeV}$.  Such a large value is extremely
unnatural given that $f_{K(1460)}/f_K\rightarrow 0$ in
the $SU(3)_F$ chiral limit.  Not surprisingly, therefore,
the lower bound $\left[ \Delta R^{(0,0)}_\tau\right]^{K(1460)}_L> 0.10$
turns out to be more than an order of magnitude larger than the value 
obtained from the sum rule analysis of the
next section.  As such, it is 
completely incompatible with the sum rules for the $us$
pseudoscalar correlator.  

\section{The Excited Strange Scalar and Pseudoscalar Resonance
Decay Constants From Scalar and Pseudoscalar Sum Rules}

From Eqs.~(\ref{wardID}) and the Ward identities for the divergences
of the flavor $ij$ V and A currents, it follows that
\begin{eqnarray}
q^4\rho^{(0)}_{ij;V}(q^2)&=&\left( m_i-m_j\right)^2\rho_{ij;S}(q^2)
\equiv \rho_{ij;\partial V}(q^2)\nonumber\\
q^4\rho^{(0)}_{ij;A}(q^2)&=&\left( m_i+m_j\right)^2\rho_{ij;P}(q^2)
\equiv \rho_{ij;\partial A}(q^2)\ ,
\label{rhorelation}
\end{eqnarray}
where $\rho_{ij;S,P}(q^2)$ is the spectral function of $\Pi_{ij;S,P}(q^2)$.
The contribution of the $K(1460)$ to $\rho^{(0)}_{us;A}$ 
on the LHS is, in the NWA,
$2f_{K(1460)}^2\, \delta\left( q^2-m^2_{K(1460)}\right)$, 
with the usual Breit-Wigner generalization to finite width.
The decay constant, $f_{K(1460)}$, is defined as usual by
\begin{equation}
\langle 0\vert A^{us}_\mu\vert K(1460)(q)\rangle =
{\rm i}\sqrt{2}f_{K(1460)}q_\mu\ .
\end{equation}
The analogous NWA contribution of the $K(1460)$ to the RHS of
Eq.~(\ref{rhorelation}) is 
$2f_{K(1460)}^2\, m^4_{K(1460)}\, \delta\left( q^2-m^2_{K(1460)}\right)$.
A similar relation holds between the $K_0^*(1430)$ contributions
to $\rho^{(0)}_{us;V}$ and $\left( m_s-m_u\right)^2 \rho_{us;S}$.
It is thus possible, in principle, to determine the
longitudinal resonance contributions of the last section indirectly, by
fixing $f_{K(1460)}$ and $f_{K_0^*(1430)}$ through analyses of the
$us$ pseudoscalar and scalar correlators, respectively.
In this section we will show that such a determination
is, indeed, feasible.  

The method involves the analysis
of the correlators of the divergences of the $us$ V and A currents, 
$\Pi_{us;\partial V}\equiv\left( m_s-m_u\right)^2\Pi_{us;S}$
and $\Pi_{us;\partial A}\equiv \left( m_s+m_u\right)^2\Pi_{us;P}$
using finite energy sum rules (FESR) of a type (``pinch-weighted'')
known to allow an accurate reconstruction of the isovector
vector spectral function using as input only the OPE, together with
PDG values for the resonance masses and
widths~\cite{kmfesr,kma0}.
We will refer to these correlators as the strange scalar (SS)
and strange pseudoscalar (SPS) correlators in what follows.

The basic idea of the analysis is straightforward.  Analyticity
leads to the general FESR relation,
\begin{equation}
\int_{s_{th}^{SS,SPS}}^{s_0}\, ds\, \rho_{us;\partial V,\partial A}(s)\, w(s)
={\frac{-1}{2\pi i}}\, \oint_{\vert s\vert =s_0}\, ds\,
\Pi_{us;\partial V,\partial A}(s)\, w(s)\ ,
\label{fesr}
\end{equation}
valid for any function $w(s)$ analytic in the region of
the contour.  The LHS is determined
by the decay constants of the relevant scalar or pseudoscalar
resonances, while the RHS, for large enough $s_0$ can be evaluated
using the OPE.  
For the case of the analogous isovector vector correlator, it has been
shown that, although the breakdown of the OPE near the 
timelike real axis for $s_0\sim m_\tau^2$ is not negligible
(so that, for example, FESR's with $w(s)=s^k$ are {\it not} 
well-satisfied for $s_0\sim m_\tau^2$~\cite{kmfesr}),
even a single zero in $w(s)$ at
$s=s_0$ is enough to produce FESR's that are very well satisfied
when the OPE representation for $\Pi_{ud;V}$ is
employed in the analogue of the RHS of Eq.~(\ref{fesr})~\cite{kmfesr}.
A physical understanding of the origin of this behavior
is provided by the arguments of Ref.~\cite{pqw}.  
As shown in Refs.~\cite{kmfesr,kma0},
working simultaneously with FESR's based on the weight families
\begin{eqnarray}
w_N(y,A)&\equiv& (1-y)(1+Ay)\nonumber\\
w_D(y,A)&\equiv& (1-y)^2(1+Ay)\ ,
\label{pinchedweights}
\end{eqnarray}
where now $y=s/s_0$, and $A$ is a free parameter, gives enough 
variability in the weight profile to strongly constrain the resonance
decay constants in a given channel{\begin{footnote}{This 
is true even for correlators for which
the spectral function contains significant background
contributions near threshold.  As an example, consider the
$us$ scalar channel.  In Ref.~\cite{cfnp}, an ansatz for the
corresponding spectral function has been constructed,
employing the Omnes representation for the timelike scalar
$K\pi$ form factor in combination with certain additional assumptions.
The resulting spectral function displays a very significant background
contribution near threshold (associated with the strongly attractive
$s$-wave $I=1/2$ $K\pi$ interaction) which cannot be well
represented by the tail of the $K_0^*(1430)$ resonance.
If one takes as input on the OPE side of the $w_N$ and
$w_D$ FESR's the value of $m_s$ obtained from a FESR analysis
using this spectral ansatz as input, and then, with the OPE
representation so fixed, 
makes an incoherent-sum-of-Breit-Wigner-resonance
ansatz for the spectral function, and uses matching to
the OPE sides of the set of $w_N$ and $w_D$ FESR's 
above to fix the resonance
decay constants, one finds that the $K_0(1430)$ peak of
the spectral function is reproduced to within $\sim 2\%$,
despite the fact that the near-threshold region is, of
course, not well-reproduced.  The reason is obvious:
the spectral function is small in the near-threshold
region and integrals over the spectral function are
sensitive dominantly to the regions where it is large, i.e., to
the regions of the resonance peaks.}\end{footnote}}.
We will denote the class of pinch-weighted FESR's
as pFESR's in what follows.

The OPE representations of the SS and SPS correlators are known up to
dimension $D=6$~\cite{jm,cdps}, the $D=0$ part being determined to
four loop order ($O(\alpha_s^3)$)~\cite{chetyrkin97}.  It 
is convenient to work with
the second derivative of $\Pi$ with respect to $Q^2$, which
satisfies a homogeneous RG equation, allowing logarithms to
be summed by the scale choice $\mu^2 =Q^2$.  One has, for the
resulting OPE representations~\cite{jm,cdps},
\begin{eqnarray}
\left[ \Pi^{\prime\prime}_{us;SPS/SS}(Q^2)\right]_{D=0}&=&
{\frac{3}{8\pi^2}}{\frac{(m_s\pm m_u)^2}{Q^2}}
\left[ 1+{\frac{11}{3}}a(Q^2)+14.1793a(Q^2)^2+77.3683a(Q^2)^3\right]
\nonumber\\
\left[ \Pi^{\prime\prime}_{us;SPS/SS}(Q^2)\right]_{D=2}&=&
{\frac{3}{4\pi^2}}{\frac{(m_s\pm m_u)^2m_s^2}{Q^4}}
\left[ 1+{\frac{28}{3}}a(Q^2)+\left( {\frac{8557}{72}}
- {\frac{77}{3}}\zeta (3)\right) a(Q^2)^2\right]
\nonumber\\
\left[ \Pi^{\prime\prime}_{us;SPS/SS}(Q^2)\right]_{D=4}&=&
{\frac{(m_s\pm m_u)^2}{Q^6}}
\Biggl[ {\mp}2\left( 1+{\frac{23}{3}}a(Q^2)\right) \langle m_s\bar{u}u\rangle
-{\frac{1}{9}}\left(1+{\frac{121}{18}}a(Q^2)\right) I_G
\Bigg. \nonumber\\
&&\Bigg. \quad 
+\left(1+{\frac{64}{9}}a(Q^2)\right) I_s -{\frac{3m_s(Q^2)^4}{7\pi^2}}
\left( {\frac{1}{a(Q^2)}}+{\frac{155}{24}}
\right)\Biggr] \nonumber\\
\left[ \Pi^{\prime\prime}_{us;SPS/SS}(Q^2)\right]_{D=6}&=&
{\frac{(m_s\pm m_u)^2}{Q^8}}\Biggl( \pm 3\left[
\langle m_ig\bar{q}_j\sigma\cdot G q_j+m_jg\bar{q}_i\sigma\cdot G q_i\rangle
\right]\nonumber \Bigg. \\
\Bigg. &&\qquad -{\frac{32}{9}}\pi^2 a\rho_{VSA}
\left[ \langle \bar{q}_i q_i\rangle^2
+\langle \bar{q}_j q_j\rangle^2 - 9\langle \bar{q}_i q_i\rangle
\langle \bar{q}_j q_j\rangle\right]\Biggr)\ ,
\label{basicope}\end{eqnarray}
where $I_G$ and $I_s$ are the RG invariant versions of the gluon
and strange quark condensate, as defined in Ref.~\cite{cps}, 
$\rho_{VSA}$ describes the deviation of the four-quark condensates
from their vacuum saturation values, and
the upper (lower) sign corresponds throughout to the SPS (SS) case.

One should bear in mind that, on the theoretical side of SS (SPS)
sum rules, the contribution of direct instantons to the SS (SPS) correlator
is not contained in the OPE representation.  
Such effects are known to play a potentially
important role in scalar and pseudoscalar
channels~\cite{novikov81,shuryak82,shuryak83,dorokhov90}, particularly at lower
scales $\sim 1\ {\rm GeV}$.{\begin{footnote}{The Borel transform of
the OPE representation of the correlator of the flavor $ij$
pseudoscalar density, for example, displays the wrong dependence 
on the Borel mass, $M$, in the
chiral limit:  while the $exp(-s/M^2)$-weighted spectral integral 
must become independent of $M$ in this limit, the Borel
transformed OPE representation displays a strong dependence on $M$.}
\end{footnote}}  As a result, one must include an estimate of
direct instanton contributions, in addition to OPE contributions,
in the theoretical representation of the SS (SPS) correlator.
A convenient, and phenomenologically constrained,
model for making such an estimate
is the instanton liquid model 
(ILM)~\cite{ilm}{\begin{footnote}{The incorrect
$M$ dependence of the theoretical side of 
the Borel transformed pseudoscalar correlator sum rule
is cured once the OPE representation is supplemented with ILM 
contributions~\cite{shuryak82,shuryak83}.}\end{footnote}}.
ILM contributions to the theoretical side of polynomial-weighted 
SPS pFESR's can be obtained from the result\cite{elias98}
\begin{equation}
{\frac{-1}{2\pi i}}\, \oint_{\vert s\vert =s_0} ds\, s^k
\left[ \Pi_{us;P}(s)\right]_{ILM} =
{\frac{-3[m_s+m_u]^2\eta_{us}}{4\pi}}\, \int_0^{s_0} ds\, s^{k+1}J_1\left(
\rho_I\sqrt{s}\right) Y_1\left(\rho_I\sqrt{s}\right)\ ,
\label{fesrinstanton}
\end{equation}
where $\rho_I\simeq (1/0.6\ {\rm GeV})$ is the average
instanton size (a parameter of the ILM),
$\eta_{us}$ is an $SU(3)$-breaking
factor whose value in the ILM is $\sim 0.6$~\cite{shuryak83},
and the result is relevant to scales $\sim 1\ {\rm GeV}^2$.
The corresponding result for the SS channel is obtained by
the replacement $(m_s+m_u)^2\rightarrow -(m_s-m_u)^2$.  

It is important to remember that, for a given scale, the
ILM contribution to a typical scalar or pseudoscalar
FESR is much larger than that
to the corresponding Borel sum rule (BSR){\begin{footnote}{The
reason is straightforward: ILM contributions to the scalar
and pseudoscalar correlators, $\Pi_{ij;S,P}(Q^2)$, 
are proportional to 
$Q^2\left[ K_{-1}\left(\rho_I \sqrt{Q^2}\right)\right]^2$.
The modulus of the MacDonald function $K_{-1}(\rho_I\sqrt{Q^2})$, 
on the circle 
$Q^2=\vert Q^2\vert e^{i\theta}$, is typically much larger for
non-zero $\theta$ than it is for the spacelike point, $\theta =0$.
The integral around the circle $\vert s\vert =s_0$ present 
on the theoretical side of a FESR thus samples regions of the complex 
$Q^2$-plane where the ILM contributions are enhanced.}\end{footnote}}.
At the scales we will be employing, ILM contributions
to the SPS and SS BSR's are, in fact, quite small, while
those to our pFESR's are still non-negligible.  
Consistency between pFESR and BSR analyses thus represents
a non-trivial constraint on the reliability
of the ILM representation of instanton contributions~\cite{kmjkpssr}.
In Ref.~\cite{kmjkpssr}, this consistency check was implemented
as follows.  First, the families of pFESR's noted above were used
to make a simultaneous determination of $m_i +m_j$ and 
the resonance decay constants relevant to the flavor $ij$ 
pseudoscalar channel, the values obtained for $m_i+m_j$
and the resonance decay constants being sensitive
to whether or not the OPE was supplemented with ILM contributions.
The resulting pFESR-generated values
for the decay constants were then used as input to a
BSR analysis of the same correlator, an alternate
determination of $m_i+m_j$ being the output of this analysis.
The pFESR and BSR determinations of $m_i+m_j$ should then
be consistent. The only non-trivial sensitivity
to ILM contributions in the BSR analysis is that associated with
the input pFESR values for the resonance decay constants.
Consistency of the two determinations was found only when ILM contributions
were included on the theoretical sides of the pFESR's~\cite{kmjkpssr}.

In what follows, therefore, we will
employ the ILM to estimate direct instanton effects, and determine the
strange scalar and pseudoscalar resonance decay constants 
in a combined  $w_N$, $w_D$ pFESR analysis.
Compatibility of the pFESR and BSR quark mass determinations 
will be imposed as an additional consistency requirement
{\begin{footnote}{Errors associated with uncertainties in
the input resonance masses and widths and the input values of
parameters appearing on the theoretical sides
of the sum rules occur for both the pFESR and BSR analyses
and are strongly correlated. 
The BSR analysis has additional errors associated with
the use of the ``continuum'' approximation for the
high-$s$ part of the spectral integral and the uncertainty
in the choice of the ``continuum threshold'' parameter.
For the SPS case, these were estimated to produce an uncertainty 
of $\sim 9\%$ in $m_s+m_u$~\cite{kmjkpssr}.  We have employed this
same estimate for the additional BSR uncertainty 
in our combined pFESR/BSR SS channel analysis.}\end{footnote}}.
The Borel transform of the ILM contribution to 
the SPS correlator required for this consistency check is given by
\begin{equation}
{\frac{3\rho_I^2\left( m_s+m_u\right)^2 M^6}{8\pi^2}}\left[
K_0(\rho_I^2 M^2/2)+K_1(\rho_I^2 M^2/2)\right] .
\label{spsbsrinst}
\end{equation}
That for the SS correlator is obtained by the replacement
$(m_s+m_u)^2\rightarrow -(m_s-m_u)^2$.
Expressions for the Borel transform of the OPE representations
are well known, and can be found in Refs.~\cite{jm,cps,dps}.

We use the following input values on the OPE+ILM side of our sum rules:
$\rho_I=1/(0.6\ {\rm GeV})$~\cite{shuryak83,ilm},
$\alpha_s(m_\tau^2)=0.334\pm .022$~\cite{ALEPHud,OPAL},
$\langle \alpha_s G^2\rangle = (0.07 \pm 0.01)\ {\rm GeV}^4$~\cite{narisonaGG},
$\left( m_u+m_d\right)\langle \bar{u}u\rangle =-f_\pi^2 m_\pi^2$ (the
GMOR relation), $0.7< \langle \bar{s}s\rangle /\langle \bar{u} u\rangle 
\equiv r_c<1$~\cite{jm,cps};
$\langle g\bar{q}\sigma Fq\rangle
=\left( 0.8\pm 0.2\ {\rm GeV}^2\right)\langle \bar{q} q\rangle$\cite{op88}
and $0<\rho_{VSA}<10$.  
The $D=0,2$ and $4$ OPE integrals are evaluated 
using the contour-improvement prescription~\cite{ledp92,pivcipt}, 
since this is known to improve convergence and reduce 
residual scale dependence~\cite{ledp92}.
The running coupling and running mass required in
this procedure are obtained using the
4-loop-truncated versions of the $\beta$~\cite{beta4} 
and $\gamma$~\cite{gamma4} functions, with the value of
$\alpha_s(m_\tau^2)$ given above as input.

The analysis of the SPS channel has already been performed in 
Ref.~\cite{kmjkpssr}, to which the interested reader
is referred for a detailed discussion.
The results of that analysis are
\begin{eqnarray}
&&m_s(2\ {\rm GeV})\, =\, 100\pm 12\ 
{\rm MeV}\label{usilmmass} \\
&&f_{K(1460)}\, =\, 21.4\pm 2.8\ {\rm MeV}\label{usilmf1} \\
&&0< f_{K(1830)}< 8.9\ {\rm MeV}\ ,
\label{usilmf2}
\end{eqnarray}
where the errors have been obtained by combining the ``theory'' and
``method'' errors of Ref.~\cite{kmjkpssr} 
in quadrature{\begin{footnote}{The method
errors refer to changes in the output produced by varying
the $s_0$ and $A$ ranges used in the pFESR analysis,
or by performing $w_N$ or $w_D$ family analyses separately,
rather than a combined analysis.  A breakdown of the
contributions to the combined error may be found in
Ref.~\cite{kmjkpssr}.}\end{footnote}}.  The lack of a strong
constraint on $f_{K(1830)}$ is a result of the smallness of
the $K(1830)$ contribution to the various pFESR spectral integrals.
Since only the tail of the $K(1830)$
contributes to hadronic $\tau$ decay, and the endpoint 
region is strongly suppressed by the kinematic weight factor,
this uncertainty plays a negligible role for our purposes.
Two further points should be stressed:
first, the value of $m_s$ obtained from the SPS pFESR/BSR analysis
is consistent with that obtained from recent analyses based
on hadronic $\tau$ decay data and, second,
even if one completely neglects ILM contributions (ignoring
the resulting inconsistency between pFESR and BSR mass determinations),
one obtains a value $f_{K(1460)}=22.9\pm 2.7\ {\rm MeV}$
compatible with that given in Eq.~(\ref{usilmf1}) within
errors{\begin{footnote}{The biggest impact of neglecting ILM
contributions is on $f_{K(1830)}$, which becomes $14.5\pm 1.4\ {\rm MeV}$.
The pFESR value of $m_s(2\ {\rm GeV})$ is also altered, the central value
becoming $116\ {\rm MeV}$, but it is difficult to assign meaningful
errors to this number since the pFESR and BSR determinations
are not consistent in this case.}\end{footnote}}.
For the purposes of determining the strange pseudoscalar 
longitudinal contributions in hadronic $\tau$ decay, the
result of Eq.~(\ref{usilmf1}) thus appears very robust.

A simultaneous pFESR determination of $m_s+m_u$ and the excited resonance
decay constants in the SPS channel is possible only because
one part of the spectral function (the $K$ pole contribution) 
is well determined experimentally.  Unfortunately the experimental 
spectral constraints are considerably weaker in the SS channel.

The SS spectral function should be dominated by contributions
from $K\pi$ intermediate states up to and including
the $K_0^*(1430)$ region, since the $K_0^*(1430)$ 
displays essentially no inelasticity.  Unitarity
and the Omnes representation of the timelike scalar $K\pi$
form factor allow one to represent the $K\pi$ component
of the spectral function in terms of $K_{\ell 3}$ 
data and $K\pi$ phases~\cite{jm,cfnp}. 
There are, however, ambiguities in this representation.
In the literature, it has been assumed that a possible polynomial prefactor
is absent from the Omnes representation and, in addition, 
that the corresponding asymptotic behavior of the $K\pi$ phase 
required by quark counting rules has already been reached at 
the upper edge of the currently
accessible experimental range, $s\simeq 2.9\ {\rm GeV}^2$.
The spectral ansatz which results from these 
assumptions~\cite{cfnp} serves as the basis for
a number of recent sum rule of analyses of the SS channel~\cite{srss}
and corresponds (reading from Fig.~2 of Ref.~\cite{cfnp})
to the constraint
\begin{equation}
26.2\ {\rm MeV}<f_{K_0^*(1430)}<31.0\ {\rm MeV}\ .
\label{cfnpfval}
\end{equation}
The corresponding value for $m_s$ is,
averaging the errors quoted in Refs.~\cite{srss}(d) and \cite{jm}(b),
\begin{equation}
m_s(2\ {\rm GeV})=115\pm 15\ {\rm MeV}\ .
\label{cfnpms}
\end{equation}
It turns out, however, that quite sizeable deviations 
from the asymptotic value of the phase are allowed
in the region of the $K_0^*(1950)$ 
without violating the known ChPT constraint
on the slope of the form factor at $s=0$.  These
can, in turn, produce non-trivial deviations of the
spectral function from that obtained in Ref.~\cite{cfnp},
even in the region of the $K_0^*(1430)$.  There are thus
potentially significant uncertainties not yet reflected in the
range of values for $f_{K_0^*(1430)}$ given in Eq.~(\ref{cfnpfval}).

Without fully constrained experimental values for the SS spectral
function, the pFESR and/or BSR analyses allow us
to determine $f_{K_0^*(1430)}$ and $f_{K_0^*(1950)}$ 
only after an input value for $(m_s-m_u)^2$ 
has been provided on the OPE+ILM side of the sum rules.  
The reason is that, at the scales employed in our analyses, those
terms in the OPE proportional to $m_s^4$ are numerically
tiny, so the OPE representation is, to a very good approximation, 
proportional to $(m_s-m_u)^2$.  Thus, once one finds an optimized
spectral ansatz for a particular value of $m_s-m_u$,
say $m_s-m_u\equiv m_0$, an equally-well-optimized ansatz 
for any other value, $m_s-m_u\equiv m_1$, can be
obtained simply by rescaling the 
fitted decay constants by $m_1/m_0$.  The pFESR
analysis thus allows only a determination of the ratios $f_i/(m_s-m_u)$.  

In our SS pFESR analysis, our spectral ansatz
consists of an incoherent sum of $K_0^*(1430)$ and
$K_0^*(1950)$ Breit-Wigner resonance forms, with PDG2000
values of the resonance masses and widths.  
We employ the same pFESR analysis window as used in
our earlier study of the SPS channel, namely $3.0\ {\rm GeV}^2
\leq s_0\leq 4.0\ {\rm GeV}^2$ and $0\leq A\leq 4$.
The different $A$ values correspond
to weights with significantly different
relative weightings between the first and second resonance
regions, and hence are useful in tightening constraints
on the resonance decay constants.  As noted above,
as a self-consistency check on the ILM representation of direct
instanton effects, we also require consistency between
the value of $m_s-m_u$ used as input to the pFESR analysis
and that obtained as output from the corresponding BSR analysis,
in which pFESR values of the resonance decay constants
are used.
The results of this determination are
\begin{eqnarray}
f_{K_0^*(1430)}&=& \left[ 22.5\pm 2.1\right] \left({\frac{m_s(2\ {\rm GeV})}
{100\ {\rm MeV}}}\right)\label{ssf1val} \\
f_{K_0^*(1950)}&=& \left[ 17.6\pm 2.0\right] \left({\frac{m_s(2\ {\rm GeV})}
{100\ {\rm MeV}}}\right)\ ,
\label{ssf2val}
\end{eqnarray}
where we have combined all sources of error in quadrature.  
If we consider the range of $m_s$ values given in Eq.~(\ref{cfnpms}),
the corresponding range of $f_{K_0^*(1430)}$ allowed by
Eq.~(\ref{ssf1val}) is
\begin{equation}
20.4\ {\rm MeV}<f_{K_0^*(1430)}<32.0\ {\rm MeV}\ ,
\label{cfnpsscheck}
\end{equation}
which turns out to be
in good agreement with that given by Eq.~(\ref{cfnpfval}).  

With the values above for the decay constants of the
SS and SPS resonances it is straightforward to
compute the expected $\tau\rightarrow K_0^*(1430) \nu_\tau$ 
branching fraction, and also the values
of the resonance contributions to 
$\left[ \Delta R^{(0,0)}_\tau\right]_L$.
Taking $83\ {\rm MeV}<m_s(2\ {\rm GeV})<130\ {\rm MeV}$~\cite{gm01rev}, 
the result of Eq.~(\ref{ssf1val})
corresponds to
\begin{equation}
0.00003<B(\tau\rightarrow K_0^*(1430) \nu_\tau )<0.00011\ ,
\label{branchingratio}
\end{equation}
and hence satisfies the constraint given by the PDG2000 upper bound.  
As expected on kinematic grounds,
the $K(1830)$ and $K_0^*(1950)$ contributions
to $\left[ \Delta R^{(0,0)}_\tau\right]_L$ are
negligible{\begin{footnote}{The 
$K_0^*(1950)$ contribution is a factor of $\sim 20$ smaller
than the $K_0^*(1430)$ contribution, the $K(1830)$
contribution a factor of $\sim 60$ smaller than the $K(1460)$
contribution.}\end{footnote}}.  The corresponding $K(1460)$ and 
$K_0^*(1430)$ contributions, which follow from Eqs.~(\ref{usilmf1})
and (\ref{ssf1val}), are{\begin{footnote}{For definiteness,
we have computed these contributions using the value 
$\vert V_{us}\vert =0.2196\pm 0.0023$ obtained
from the analysis of $K_{e3}$ decay data.
The results, of course, scale as $1/\vert V_{us}\vert^2$.}\end{footnote}}
\begin{equation}
\left[ \Delta R^{(0,0)}_\tau\right]^{K(1460)}_L = 
0.0052\pm 0.0014
\label{k1460longcontribution}
\end{equation}
and
\begin{equation}
\left[ \Delta R^{(0,0)}_\tau\right]^{K_0^*(1430)}_L = 
\left[ 0.0059\pm 0.0011\right] \left[ {\frac{m_s(2\ {\rm GeV})}
{100\ {\rm MeV}}}\right]^2\ .
\label{k1430longcontribution}
\end{equation}
The sum of the SS and SPS resonance contributions is thus $\sim 10\%$
of the $K+\pi$ pole contribution.  This level of suppression of resonance
relative to pole contributions is in fact quite natural, and
represents a combination of chiral and kinematic effects.
With $y_K\equiv m_K^2/m_\tau^s$ and 
$y_{res}\simeq \left( 1.4\ {\rm GeV}^2\right) /m_\tau^2$,
one has $w_L^{(0)}(y_K)=.13$ and $w_L^{(0)}(y_{res})=.16$.  Thus,
although $w_L^{(0)}(y_K)$, which is proportional to
$m_K^2$, and hence of $O(m_s)$ in the chiral
counting, is formally suppressed by one power of $m_s$ relative
to $w_L^{(0)}(y_{res})$, the factor $(1-y)^2$ 
in $w_L^{(0)}(y)$ produces a kinematic suppression of
$w_L^{(0)}(y_{res})$ which largely undoes this effect.
As a result, one expects resonance contributions to
$\left[ \Delta R^{(0,0)}_\tau\right]_L$ to be suppressed,
relative to the $K$ contribution, by 
the ratio $\left[ f_{K_0^*(1430),K(1460)}/f_K\right]^2$,
which has an $m_s^2$ chiral suppression, $\sim 0.1$.  Naive
chiral counting would have produced instead a less
strong suppression, of order $m_s$, $\sim 0.3$.

As noted above, the result of Eq.~(\ref{k1460longcontribution})
is more than an order of magnitude smaller than the lower bound implied
by the combination of the PDG2000 upper bound on the $K_0^*(1430)$
branching fraction and the assumption that the OPE 
representation of the longitudinal $(0,0)$ spectral weight
contribution is reliable.  To satisfy the lower
bound, one would require a value of $f_{K(1460)}$
a factor of $\sim \sqrt{20}$ larger than that given above.
Such a large value, however, 
leads to an exceptionally poor ``optimized'' pFESR OPE/spectral match.
The value of $m_s$ corresponding to this ``optimized''
match, moreover, produces a $(0+1)$ OPE contribution
which already exceeds the experimental value for
$\Delta R^{(0,0)}_\tau$, and hence violates the positivity of
the longitudinal SS and SPS contributions.  

\section{Summary and Discussion}
We have shown that determinations of $m_s$
based on inclusive $(k,0)$ spectral weight analyses of
flavor breaking in hadronic $\tau$ decay have an unavoidable,
unphysical dependence on $k$, and that the impact
of this unphysical behavior on the extracted values for
$m_s$ is numerically significant.  The problem 
has been shown to result from the unphysical behavior with respect to $k$
of the OPE representation of the longitudinal contributions
to the $ud$ and $us$ correlators, when truncated at $D=6$.
If truncation at $D=6$ is justified, then the problem
lies with the $D=2$ part of the OPE representation
(whose integrated contour-improved series, in any
case, already displays rather bad non-converging behavior).  
Part of the problem, however, may lie in the neglect of higher
dimension contributions, particularly since contributions
un-suppressed by additional factors
of $\alpha_s$, and having dimensions up to $D=8,10$ and $12$, 
are in principle present for the
$(0,0)$, $(1,0)$ and $(2,0)$ analyses, respectively.
Since nothing is known, phenomenologically,
about the values of $D=8$ and higher condensates, the
only way to investigate this question is to 
consider spectral weight (or other pFESR) analyses 
with $s_0\not= m_\tau^2$, and try to use the 
$s_0$ dependence to separate contributions
of different dimension{\begin{footnote}{Contributions of
dimension $D=2N$ scale with $s_0$ as $1/s_0^{N-1}$, up
to logarithms.}\end{footnote}}.  To work with $s_0\not = m_\tau^2$,
however, one must necessarily perform a non-inclusive analysis,
since different kinematic factors, both of which are specific to
$s_0=m_\tau^2$, are associated
with the $0+1$ and longitudinal contributions to the
experimental number distribution.
In the region of the spectrum where
the separation into $0+1$ and longitudinal
contributions is not straightforward (the excited SS and
SPS resonance region) we have provided determinations
of the SS and SPS resonance decay constants accurate to $\sim 10\%$.
This allows one to evaluate the resonance part of the longitudinal
contribution with an accuracy of $\sim 20\%$.  
Even for the $(0,0)$ analysis, such an uncertainty corresponds
to only a $\sim 2\%$ uncertainty on the total longitudinal
subtraction; for weights which more strongly suppress the excited resonance
region, the corresponding uncertainty is even smaller
{\begin{footnote}{Although the large amount of new $\tau$ 
data that will be generated by the B factory experiments
will eventually dramatically change the experimental situation,
at present, experimental errors on the $V+A$ $us$ number
distribution are quite large ($\sim 20-30\%$) beyond the $K^*$
region.  As a result, reduced errors on $m_s$, at least at present,
require use of weights which fall off with $s$ in this region
more strongly than does the $(0,0)$ transverse weight.
The uncertainties on our determinations of the decay constants
thus play a negligible role in current analyses.}\end{footnote}}.

We conclude with a comment on the implications of our results
for future $\tau$ decay determinations of $m_s$.
It is our opinion that the bad behavior of the
OPE representation of the longitudinal contributions precludes
the reliable use of an inclusive analysis, and forces us to
make a theoretical evaluation of the longitudinal resonance
contributions to the spectrum.  
As a result, the previous dis-incentive to studying 
the $s_0$ dependence of any particular pFESR
(the non-inclusivity of such an analysis) is no longer in play.  
Since the $D=2,4,6$ terms in the OPE representation 
of $\Delta\Pi^{(0+1)}$ are well-behaved,
a study of the $s_0$ dependence then becomes
crucial either to demonstrating explicitly that higher dimension
contributions can, indeed, be safely neglected or 
to constraining their magnitude, if they cannot.
Because of the very strong correlations between spectral
integrals corresponding to different $s_0$, but fixed weight, 
$w(s/s_0)$, truncated OPE representations which either
miss, or pass obliquely through the experimental error band 
for the $s_0$-dependent spectral integral results, will
both signal the presence of such neglected higher $D$ terms.
Our determinations of the SS and SPS resonance decay constants
make it possible for the $s_0$ dependence of the $0+1$ sum rules 
to be studied in a straightforward manner, and we believe
that such a study should be part of all future investigations.

\acknowledgements
{KM acknowledges the ongoing support of the Natural Sciences and
Engineering Research Council of Canada, the hospitality of the
Special Research Centre for the Subatomic Structure of Matter at the
University of Adelaide, and the Theory Group at TRIUMF,
and useful conversations with Alexei Pivovarov.
JK acknowledges the partial support of the Schweizerischer Nationalfonds and
EC-Contract No. ERBFMRX-CT980169 (EURODA$\Phi$NE).}
\vskip .5in

\noindent
\begin{table}[htbp]
\caption{Comparison of longitudinal and total OPE contributions
to $\Delta R^{(k,0)}_\tau$, as obtained in Ref.~[10].
The quoted values of $m_s(m_\tau^2)$ are those obtained
in Ref.~[10] using the given spectral
weight, $(k,0)$.  This means that the difference between
the total longitudinal OPE contribution and the experimental
value, $\left[ \Delta R^{(k,0)}_\tau\right]^{exp}$, 
is the value of the $0+1$ OPE contribution produced by the
given $m_s(m_\tau^2)$.  The decrease of the longitudinal
$D=2$ contributions with $k$ is a reflection of the
decrease in the extracted $m_s(m_\tau^2)$ with $k$; as explained in the
text, for a fixed $m_s(m_\tau^2)$, the longitudinal
$D=2$ values would be increasing with $k$. The experimental
errors are those of the ALEPH collaboration.}
\vskip .15in \noindent
\begin{tabular}{ccccc}
Weight&$m_s(m_\tau^2)\ [{\rm MeV}]$&
$\left[ \Delta R^{(k,0)}_\tau\right]^{D=2}_L$&
$\left[ \Delta R^{(k,0)}_\tau\right]^{OPE}_L$&
$\left[ \Delta R^{(k,0)}_\tau\right]^{exp}$ \\
\hline
$(0,0)$&$143$&$0.201\pm 0.085$&$0.274\pm 0.087$&$0.394\pm 0.137$ \\
$(1,0)$&$121$&$0.150\pm 0.073$&$0.223\pm 0.076$&$0.383\pm 0.078$ \\
$(2,0)$&$106$&$0.126\pm 0.072$&$0.199\pm 0.074$&$0.373\pm 0.054$ \\
\end{tabular}
\label{table1}
\end{table}

\noindent
\begin{table}[htbp]
\caption{Longitudinal $\pi$+$K$ pole contributions to
$\left[ \Delta R^{(k,0)}_\tau\right]_L$, together with
the resonance contributions implicit in the longitudinal
OPE representations of Ref.~[10].  The column labelled
'IND' gives the latter results with each
$\left[ \Delta R^{(k,0)}_\tau\right]^{OPE}_L$
evaluated using the central value of the corresponding
independent fit for $m_s(m_\tau^2)$.  The column labelled
'COMB' gives the same results, except that now the
$\left[ \Delta R^{(k,0)}_\tau\right]^{OPE}_L$ are all
evaluated using the common value $m_s(m_\tau^2)=119\ {\rm MeV}$
obtained in the combined analysis using the $(0,0)$, $(1,0)$,
and $(2,0)$ spectral weights.}
\vskip .15in\noindent
\begin{tabular}{cccc}
Weight&$\left[ \Delta R^{(k,0)}_\tau\right]^{K+\pi}_L$&
$\left[ \Delta R^{(k,0)}_\tau\right]^{resonance}_L$ (IND)&
$\left[ \Delta R^{(k,0)}_\tau\right]^{resonance}_L$ (COMB) \\
\hline
$(0,0)$&$0.1204$&$0.154\pm 0.087$&$0.092\pm 0.062$ \\
$(1,0)$&$0.1105$&$0.112\pm 0.076$&$0.107\pm 0.074$ \\
$(2,0)$&$0.1014$&$0.097\pm 0.074$&$0.130\pm 0.092$ \\
\end{tabular}\label{table2}
\end{table}

\vfill\eject

\vfill\eject

\begin{references}
\bibitem{srss}(a) K.G. Chetyrkin, D. Pirjol and K. Schilcher, 
Phys. Lett. {\bf B404}, 337 (1997); (b) T. Bhattacharya, 
R. Gupta and K. Maltman, Phys. Rev. {\bf D57}, 5455 (1998);
(c) P. Colangelo, F. De Fazio, G. Nardulli and N. Paver, 
Phys. Lett. {\bf B408}, 340 (1997);
(d) K. Maltman, Phys. Lett. {\bf B462}, 195 (1999).
\bibitem{jm}(a) M. Jamin and M. M\"unz, Z. Phys. {\bf C66}, 633 (1995);
(b) M. Jamin, Nucl. Phys. B (Proc. Suppl.) {\bf 64}, 250 (1998).
\bibitem{sr38}S. Narison, Phys. Lett. {\bf B358}, 113 (1995);
K. Maltman, Phys. Lett. {\bf B428}, 179 (1998);
S. Narison, Phys. Lett. {\bf B466}, 345 (1999); hep-ph/9905264.
\bibitem{srsps}C. Dominguez, L. Pirovano and K. Schilcher, Phys. Lett.
{\bf B425}, 193 (1998).
\bibitem{ck93}K.G. Chetyrkin and A. Kwiatkowski, Z. Phys. {\bf C59}, 
525 (1993) and hep-ph/9805232.
\bibitem{kmtpr}K. Maltman, Phys. Rev. {\bf D58}, 093015 (1998).
\bibitem{pp98}A. Pich and J. Prades, JHEP {\bf 9806}, 013 (1998).
\bibitem{cdh}S. Chen, M. Davier and A. Hocker, LAL-98-90, Nov. 1998.
\bibitem{ckp98}K.G. Chetyrkin, J.H. Kuhn and A.A. Pivovarov,
Nucl. Phys. {\bf B533}, 473 (1998).
\bibitem{pp99}A. Pich and J. Prades, JHEP {\bf 9910}, 004 (1999).
\bibitem{kkp00}J.G. K\"orner, F. Krajewski and A.A. Pivovarov, hep-ph/0003165.
\bibitem{km00}J. Kambor and K. Maltman,  Phys. Rev. {\bf D62}, 093023 (2000);
Nucl. Phys. Proc. Suppl. {\bf 98}, 314 (2001).
\bibitem{D00}M. Davier, S.M. Chen, A. H\"ocker, J. Prades and
A. Pich, Nucl. Phys. Proc. Suppl. {\bf 98}, 319 (2001);
see also A. Pich, Nucl. Phys. Proc. Suppl. {\bf 98}, 385 (2001). 
\bibitem{C00}S. Chen {\it et al.}, hep-ph/0105253.
\bibitem{lat1}R. Gupta and T. Bhattacharya, Phys. Rev. {\bf D55}, 7203
(1997), T. Bhattacharya and R. Gupta, Nucl. Phys. (Proc. Suppl.) 
{\bf 63}, 95 (1998);
B.J. Gough {\it et al.}, Phys. Rev. Lett. {\bf 79}, 1662 (1997).
\bibitem{lat2}R.D. Kenway, Nucl. Phys. (Proc. Suppl.) {\bf 73}, 16 (1999).
\bibitem{lat3}V. Lubicz, Nucl. Phys. (Proc. Suppl.) {\bf 74}, 291 (1999).
\bibitem{latticerecent}T. Blum, A. Soni and M. Wingate, Phys. Rev. {\bf D60}, 
114507 (1999);
J. Gardner, J. Heitger, R. Sommer and H. Wettig
(ALPHA/UKQCD), Nucl. Phys. {\bf B571}, 237 (2000);
S. Aoki {\it et al.} (JLQCD), Phys. Rev. 
Lett. {\bf 82}, 4392 (1999);
S. Aoki {\it et al.} (CP-PACS), Phys. Rev.
Lett. {\bf 84}, 238 (2000);
W. G\"ockeler {\it et al.} (QCDSF), Phys. Rev. {\bf D62}, 054504 (2000);
D. Becirevic, V. Lubicz, G. Martinelli
and M. Testa, Nucl. Phys. (Proc. Suppl.) {\bf 83}, 863 (2000);
A. Ali Khan {\it et al.} (CP-PACS), Nucl. Phys. (Proc. Suppl.) {\bf 83},
176 (2000) and Phys. Rev. Lett. {\bf 85}, 4674 (2000).
\bibitem{gm01rev}R. Gupta and K. Maltman, hep-ph/0101132.
\bibitem{ALEPHud}R. Barate {\it et al.} (The ALEPH Collaboration),
Z. Phys. {\bf C76}, 379 (1997); Eur. Phys. J. {\bf C4}, 409 (1998).
\bibitem{ALEPHstrange}R. Barate {\it et al.} (The ALEPH Collaboration),
Eur. Phys. J. {\bf C11}, 599 (1999).
\bibitem{tau1}Y.S. Tsai, Phys. Rev. {\bf D4}, 2821 (1971);
H.B. Thacker and J.J. Sakurai, Phys. Lett. {\bf B36}, 103 (1971);
F.J. Gilman and D.H. Miller, Phys. Rev. {\bf D17}, 1846 (1978);
F.J. Gilman and S.H. Rhie, Phys. Rev. {\bf D31}, 1066 (1985).
\bibitem{tau2}E. Braaten, Phys. Rev. Lett. {\bf 60}, 1606 (1988);
S. Narison and A. Pich, Phys. Lett. {\bf B211}, 183 (1988); E. Braaten, Phys.
Rev. {\bf D39}, 1458 (1989); S. Narison and A. Pich, Phys. Lett. {\bf B304},
359 (1993).
\bibitem{bnp}E. Braaten, S. Narison and A. Pich, Nucl. Phys. {\bf B373},
581 (1992).
\bibitem{ms88}W.J. Marciano and A. Sirlin, Phys. Rev. Lett. {\bf 61},
1815 (1988).
\bibitem{ledp92}F. Le Diberder and A. Pich, Phys. Lett. {\bf B286}, 147
(1992) and {\bf B289}, 165 (1992).
\bibitem{pichrev}A. Pich, hep-ph/9704453, in ``Heavy Flavors II'',
eds. A.J. Buras and M. Lindner, World Scientific, 1997.
\bibitem{chetyrkin97}K.G. Chetyrkin, Phys. Lett. {\bf B390}, 309 (1997).
\bibitem{cps}K.G. Chetyrkin, D. Pirjol and K. Schilcher,  Phys.
Lett. {\bf B404}, 337 (1997).
\bibitem{kkp001}J.G. K\"orner, F. Krajewski and A.A. Pivovarov,
Eur. Phys. J. {\bf C14}, 123 (2000).
\bibitem{pivcipt}A.A. Pivovarov, Sov. J. Nucl. Phys. {\bf 54}, 676 (1991)
and Z. Phys. {\bf C53}, 461 (1992).
\bibitem{pdg2000}Review of Particle Properties, D.E. Groom {\it et al.},
Eur. Phys. J {\bf C15}, 1 (2000). 
\bibitem{leutwylermq}H. Leutwyler, Phys. Lett. {\bf B374}, 163 (1996);
Phys. Lett. {\bf B378}, 313 (1996) and hep-ph/0011044.
\bibitem{CGL01}G. Colangelo, J. Gasser and H. Leutwyler, 
Phys. Rev. Lett. {\bf 86}, 5008 (2001). 
\bibitem{kmfesr}K. Maltman, Phys. Lett. {\bf B440}, 367 (1998).
\bibitem{kma0}K. Maltman, Phys. Lett. {\bf B462}, 14 (1999).
\bibitem{pqw}E. Poggio, H. Quinn and S. Weinberg, Phys. Rev. {\bf D13},
1958 (1976).
\bibitem{cfnp}P. Colangelo, F. De Fazio, G. Nardulli and N. Paver,
Phys. Lett. {\bf B408}, 340 (1997).
\bibitem{cdps}K.G. Chetyrkin, C.A. Dominguez, D. Pirjol and K. Schilcher,
Phys. Rev. {\bf D51}, 5090 (1995).
\bibitem{novikov81}V.A. Novikov, {\it et al.},  Nucl. Phys.
{\bf B191}, 301 (1981).
\bibitem{shuryak82}E.V. Shuryak, Nucl. Phys. {\bf B203}, 93,116 (1982).
\bibitem{shuryak83}E.V. Shuryak, Nucl. Phys. {\bf B214}, 237 (1983).
\bibitem{dorokhov90}A.E. Dorokhov and N.I. Kochelev, Z. Phys. {\bf C46},
281 (1990).
\bibitem{ilm}E.V. Shuryak, Phys. Rep. {\bf 115}, 151 (1984);
E.V. Shuryak and J.J.M. Verbaarschot, Nucl. Phys. {\bf B410}, 55 (1993);
E.V. Shuryak, Rev. Mod. Phys. {\bf 65}, 1 (1993);
T. Schafer and E.V. Shuryak,  Rev. Mod. Phys. {\bf 70}, 323 (1998).
\bibitem{elias98}A.E. Dorokhov, S.V. Esaibegian, N.I. Kochelev and N.G. 
Stefanis, J. Phys. {\bf G23}, 643 (1997);
V. Elias, F. Shi and T.G. Steele, J. Phys. G {\bf 24}, 267
(1998).
\bibitem{kmjkpssr}K. Maltman and J. Kambor, ``Decay Constants,
Light Quark Masses, and Quark Mass Bounds from Light Quark Pseudoscalar
Sum Rules'', in preparation.
\bibitem{dps}C.A. Dominguez, L. Pirovano and K. Schilcher, 
Phys. Lett. {\bf B425}, 193 (1998).
\bibitem{OPAL}G. Abbiendi {\it et al.} (The OPAL Collaboration), 
Phys. Lett. {\bf B447}, 134 (1999); Eur. Phys. J. {\bf C7}, 571 (1999).
\bibitem{narisonaGG}S. Narison, Nucl. Phys. Proc. Supp. {\bf 54A}, 238
(1997).
\bibitem{op88}A.A. Ovchinnikov and A.A. Pivovarov, Sov. J. Nucl. Phys.
{\bf 48}, 721 (1988).
\bibitem{beta4}T. van Ritbergen, J.A.M. Vermaseren and S.A. Larin,
Phys. Lett. {\bf B400}, 379 (1997).
\bibitem{gamma4}K.G. Chetyrkin, Phys. Lett. {\bf B404}, 161 (1997);
T. Van Ritbergen, J.A.M. Vermaseren and S.A. Larin,
Phys. Lett. {\bf B405}, 327 (1997).
\end{references}
\end{document}